\documentclass[letterpaper]{article} 
\usepackage{amsfonts}
\usepackage{aaai2026}  
\usepackage{times}  
\usepackage{helvet}  
\usepackage{courier}  
\usepackage[hyphens]{url}  
\usepackage{graphicx} 
\usepackage{natbib}  
\usepackage{caption} 
\usepackage{algorithm}
\usepackage{algorithmic}
\usepackage{amsmath} 
\usepackage{subcaption}
\usepackage{multirow}
\usepackage{booktabs} 
\usepackage{newfloat}
\usepackage{listings}
\DeclareCaptionStyle{ruled}{labelfont=normalfont,labelsep=colon,strut=off} 
\floatstyle{ruled}
\newfloat{listing}{tb}{lst}{}
\floatname{listing}{Listing}
\nocopyright

\title{FDC-Net: Rethinking the association between EEG artifact removal and multi-dimensional affective computing}
\author{
    \textbf{Wenjia Dong}, \textbf{Xueyuan Xu*}, \textbf{Tianze Yu}, \textbf{Junming Zhang}, \textbf{Li Zhuo} \\
    {\normalfont School of Information Science and Technology, Beijing University of Technology, Beijing 100124, China} \\
     {\normalfont \{23027425, 23027418, junming\}@emails.bjut.edu.cn, \{xxy, zhuoli\}@bjut.edu.cn}
}

\affiliations{
    \textsuperscript{\rm}

}

\usepackage{bibentry}

\begin{document}

\maketitle

\begin{abstract}
Electroencephalogram (EEG)-based emotion recognition holds significant value in affective computing and brain-computer interfaces. However, in practical applications, EEG recordings are susceptible to the effects of various physiological artifacts. Current approaches typically treat denoising and emotion recognition as independent tasks using cascaded architectures, which not only leads to error accumulation, but also fails to exploit potential synergies between these tasks. Moreover, conventional EEG-based emotion recognition models often rely on the idealized assumption of "perfectly denoised data", lacking a systematic design for noise robustness. To address these challenges, a novel framework that deeply couples denoising and emotion recognition tasks is proposed for end-to-end noise-robust emotion recognition, termed as \textbf{F}eedback-\textbf{D}riven \textbf{C}ollaborative \textbf{Net}work for Denoising-Classification Nexus (FDC-Net). Our primary innovation lies in establishing a dynamic collaborative mechanism between artifact removal and emotion recognition through: (1) bidirectional gradient propagation with joint optimization strategies; (2) a gated attention mechanism integrated with frequency-adaptive Transformer using learnable band-position encoding. Two most popular EEG-based emotion datasets (DEAP and DREAMER) with multi-dimensional emotional labels were employed to compare the artifact removal and emotion recognition performance between FDC-Net and nine state-of-the-art methods. In terms of the denoising task, FDC-Net obtains a maximum correlation coefficient (CC) value of 96.30\% on DEAP and a maximum CC value of 90.31\% on DREAMER. In terms of the emotion recognition task under physiological artifact interference, FDC-Net achieves emotion recognition accuracies of 82.3±7.1\% on DEAP and 88.1±0.8\% on DREAMER. 
\end{abstract}

\section{Introduction}
Emotions are key drivers of human cognition and behavior, influencing not only decision-making and social interactions but also being closely linked to mental health\citep{davidson2003affective}. Since traditional emotion assessment methods rely on subjective reports, they are susceptible to individual differences in expression and social desirability biases\citep{paulhus1991measurement}. 

\begin{figure}[htbp]
    \centering
 
    \begin{subfigure}[b]{1\linewidth} 
        \centering
        \includegraphics[width=\linewidth]{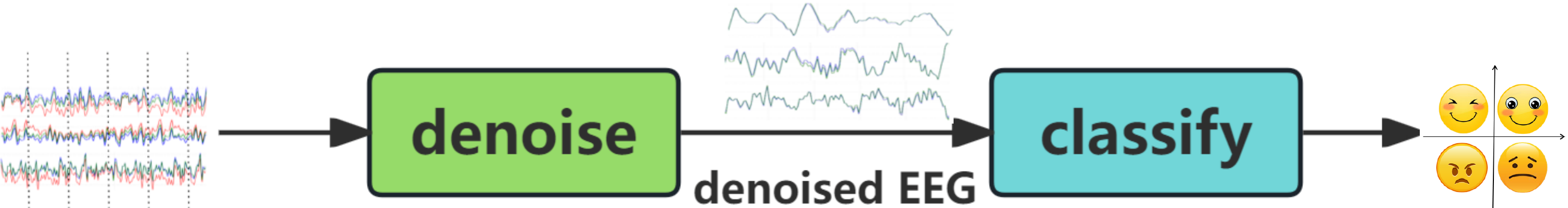}
        \caption{}          
        \label{traditional_framework}   
    \end{subfigure}
    
    \vspace {15pt}

    \begin{subfigure}[b]{1\linewidth}
        \centering
        \includegraphics[width=\linewidth]{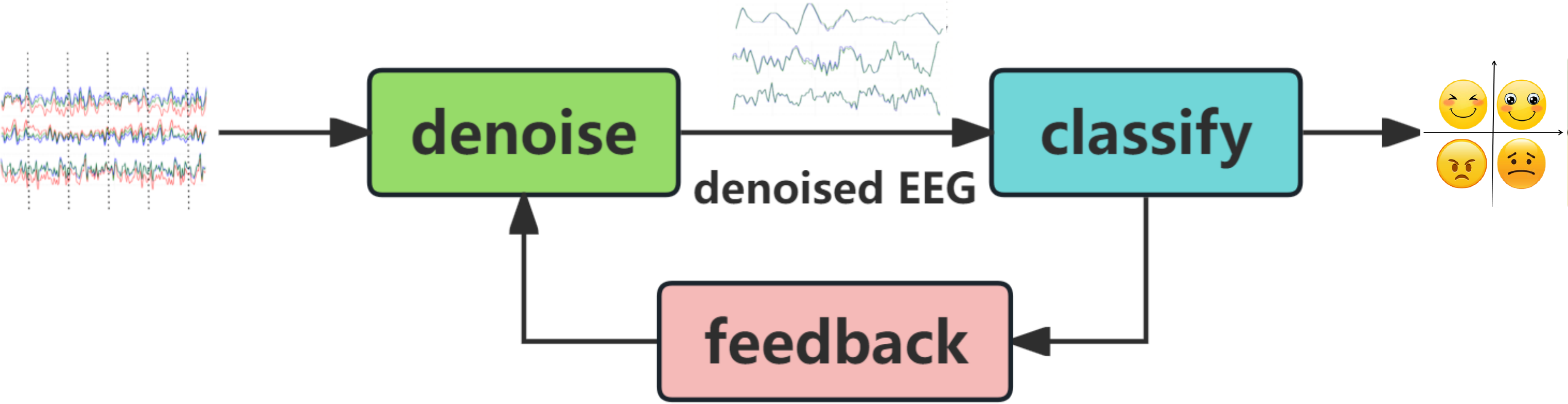}
        \caption{}          
        \label{our_framework}           
    \end{subfigure}
    
    \caption{Comparison of frameworks: (a) Existing EEG decoding approaches involve the two tasks of denoising and classification operating independently; (b) Our framework enables denoising and classification to work collaboratively after forming a feedback loop.}  
    \label{fig:framework_comparison}       
\end{figure}

In contrast, Electroencephalogram (EEG)-based emotion recognition technology, with its objectivity, high temporal resolution, and sensitivity to emotional states, has gradually become a research focus in affective computing and brain-computer interface (BCI) fields. This technology has been successfully applied in auxiliary diagnosis of depression\citep{de2019depression}, development of intelligent human-computer interaction systems \citep{frey2013review}, and monitoring of drivers' emotional states \citep{gamage2022emotion}, providing an objective and reliable physiological basis for mental health assessment and affective computing \citep{wang2022systematic}.

However, practically acquired EEG recordings inevitably suffer from contamination by various physiological artifacts, such as Electromyography (EMG), Electrooculography (EOG), and motion artifacts\citep{shad2020impedance}. These noises overlap with emotion-related neural activities in both frequency and time domains, significantly reducing the signal-to-noise ratio of raw signals. Existing EEG denoising methods, which mainly rely on techniques like independent component analysis\citep{albera2012ica}, wavelet transform \citep{adeli2003analysis}], and adaptive filtering\citep{kher2016adaptive}, could separate noise components from clean ones to a certain extent. However, these methods often require manual intervention and parameter adjustment, making them difficult to adapt to inter-individual variability. Recently, with the development of deep learning, several end-to-end denoising methods based on convolutional neural networks (CNN) and recurrent neural networks (RNN) have emerged, enabling automatic learning of complex mapping relationships between artifacts and effective signals, thus providing new solutions for EEG signal processing.

Currently, deep learning-based EEG denoising methods are mainly categorized into three types: (1) CNN was employed to capture local time-frequency features and performs well in end-to-end denoising but may ignore long-range dependencies\citep{sun2020novel}; (2) RNN was utilized to model temporal dependencies but faces gradient vanishing issues when processing long sequences\citep{yang2020decoding}; (3) and Transformers could achieve global context modeling through self-attention mechanisms\citep{chen2023denosieformer}, showing advantages in joint denoising and classification tasks but with high computational complexity. Additionally, in terms of emotion recognition models, hybrid architectures (e.g., CNN-RNN or CNN-Transformer) have become mainstream\citep{zamani2021emotion}. By combining local feature extraction and temporal modeling capabilities, they could achieve state-of-the-art performance in the EEG-based emotion recognition task\citep{yao2024emotion}. Nevertheless, these methods mostly optimize denoising and classification modules independently, failing to fully exploit the synergistic effects between dual tasks, and there remains room for improvement in noise robustness.

To address these key challenges, as shown in Fig.~\ref{fig:framework_comparison}, this paper proposes a novel \textbf{F}eedback-\textbf{D}riven \textbf{C}ollaborative \textbf{Net}work for Denoising-Classification Nexus (FDC-Net). Its core innovation is deep coupling denoising and emotion recognition tasks through a bidirectional feedback learning mechanism, enabling end-to-end noise-robust emotion recognition. Specifically, FDC-Net uses a bidirectional gradient propagation and joint optimization strategy: the denoising module adaptively retains discriminative features based on emotional semantic information fed back by the emotion recognition task, while the classification module adjusts its robustness to noise via classification loss. Notably, we also design a frequency-adaptive Transformer with learnable frequency-band positional encoding (EEGSPTransformer), which optimizes feature weights of different frequency bands in the joint task via feedback. Experimental results show that under severe noise interference, FDC-Net outperforms traditional cascaded methods in multi-dimensional emotion recognition performance. This work provides a new paradigm for EEG analysis in noisy environments, and its collaborative framework could extend to other physiological signal processing fields.

The main contributions of this paper are as follows:

\begin{itemize}
\item A dynamic feedback-driven collaborative learning framework for robust EEG-based emotion recognition is proposed. To the best of our knowledge, this is the first study to design a collaborative mechanism between EEG artifact removal and multi-dimensional emotion recognition tasks to achieve interaction via feedback. Through a joint optimization strategy, mutual promotion and performance improvement of the dual tasks are enabled.
\item A novel EEG-specific Transformer (EEGSPTransformer) module is designed. By virtue of learnable frequency-band positional encoding and an adaptive spectral attention mechanism, the time-frequency information of EEG signals are effectively captured, and the robustness of FDC-Net to artifacts is significantly enhanced by this module.
\item Comprehensive experimental validation is conducted on two most popular EEG datasets (DEAP and DREAMER) with multi-dimensional emotional labels. Compared with nine advanced methods, the experimental results of EEG artifact removal and multi-dimensional emotion recognition tasks show the superiority and robustness of the proposed method under various noise conditions.
\end{itemize}

\section{Related Works}
\subsection{EEG Denoising Research}
In recent years, deep learning-based approaches for EEG denoising have witnessed significant advancements. For example, Li et al. proposed DeepFilterNet\citep{schroter2022deepfilternet} that employs multi-scale convolutional kernels to effectively capture time-frequency representations. Yin et al. introduced a novel hybrid architecture GCTNet that integrates CNNs for local temporal feature extraction with Transformers to model long-range dependencies\citep{wang2024gctnet}. Expanding on this line of research, Cai et al. proposed a dual-branch network DHCT-GAN wherein one branch leverages a CNN-Transformer module for time-domain processing and the other utilizes a graph convolutional network (GCN) to model inter-channel topological dependencies \citep{cai2025dhct}. A key innovation lies in the introduction of a spectral-consistency discriminator, which enforces alignment between the power spectral characteristics of the denoised and clean EEG signals. Despite their effectiveness, these methods predominantly optimize for SNR or reconstruction errors, often overlooking the utility of denoised signals in downstream cognitive or affective computing tasks.

\begin{figure*}
    \centering
    \begin{subfigure}[b]{\linewidth} 
        \centering
        \includegraphics[width=\linewidth]{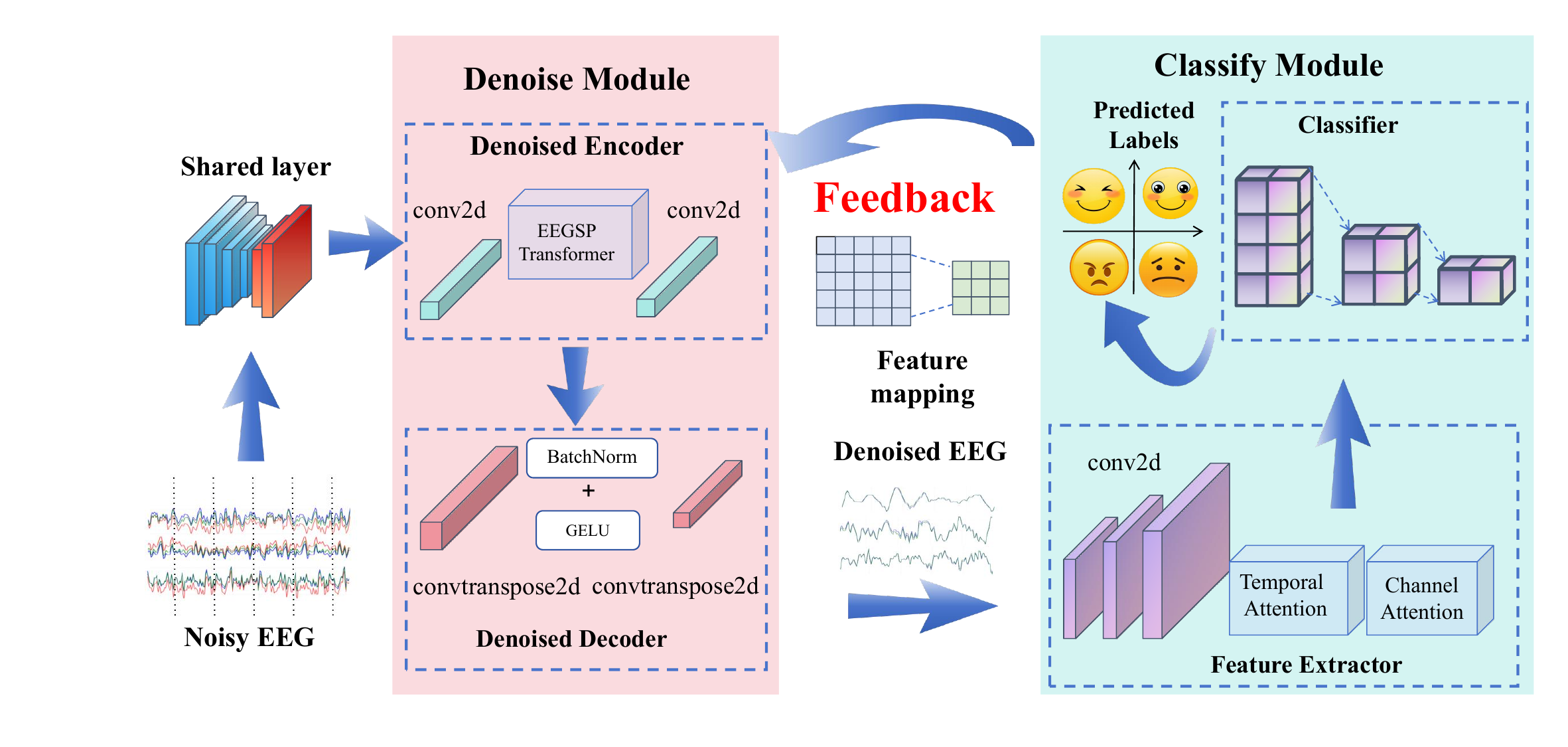}
        \label{model}
    \end{subfigure}
    \caption{The FDC-Net framework: (1) The denoising module processes noisy EEG using an encoder with an EEGSP Transformer and a decoder combining transposed convolutions, BatchNorm, and GELU; (2) The classify module extracts features for classification via convolutions and temporal-spatial attention; (3) A shared layer and feedback mechanism interact to enable dynamic collaboration between the two tasks, enhancing noise robustness.}
    \label{fig:model}
\end{figure*}

\subsection{EEG Emotion Recognition Research}
In the domain of EEG-based emotion recognition, spatio-temporal feature modeling has become a central research focus. Lu et al. presented CIT-EmotionNet\citep{lu2023cit}, a parallel CNN-Transformer architecture in which the CNN branch targets low-frequency emotional features, while the Transformer branch captures cross-frequency interactions. Devarajan et al. introduced a dynamic graph convolutional module \citep{ye2022hierarchical} with a learnable adjacency matrix, effectively capturing inter-subject variability. Wang et al. proposed DynEmoNet, a hierarchical dynamic graph convolutional framework that models both local electrode clusters and global brain-region interactions. The model incorporates a Temporal Graph Fluctuation Constraint to stabilize emotion-relevant connectivity patterns, resulting in improved emotion recognition performance. However, a common limitation of these methods is the assumption of clean pre-processed EEG inputs, which significantly undermines their robustness and generalizability in noisy environments.

In summary, existing methods tend to treat denoising and emotion recognition as independent objectives, neglecting their potential interplay. Such decoupling can inadvertently lead to the removal of emotion-relevant neural components (e.g., gamma-band oscillations) during denoising, thereby impairing emotion recognition performance when noise is present. To bridge this gap, we propose an end-to-end joint learning framework that explicitly integrates a feature-interaction attention mechanism and simultaneously optimizes a multi-task loss function. FDC-Net synergistically enhances both artifact suppression and emotion recognition performance, offering a robust and practical solution for EEG-based affective computing in real-world settings.

\section{Methodology}
We propose FDC-Net to synergistically handle EEG artifact removal and emotion recognition tasks through dual-task feedback interaction. Let $X = \{x_i\}_{i=1}^N$ denotes the raw EEG signals where each sample $x_i \in \mathbb{R}^{ C \times T}$ contains $C$ channels and $T$ time points. The model takes noisy EEG signals $x_{\mathrm{noisy}} \in \mathbb{R}^{C \times 1 \times T}$ as input, and outputs denoised signals $\hat{x}_{\mathrm{denoised}}$ and emotion labels $\hat{y} \in \{0,1\}^{2}$ (valence and arousal). The FDC-Net model comprises three components: the denoise module, the classify module, and the interaction of the dual path feature with feedback. Fig.~\ref{fig:model} illustrates the FDC-Net architecture. Each module will be elaborated on in detail as follows.

\subsection{Denoise Module}

Transformer-based methods show promise for EEG denoising but face challenges like mismatched positional encoding, inadequate spatial modeling, and underdeveloped time-frequency collaboration. These limitations drive our proposal of an EEG-specific Transformer module (\textbf{EEGSP Transformer}) with innovative designs to achieve breakthroughs. Below is a detailed introduction to this module.

\subsubsection{1. Band-Limited PE:}
Based on the constraints of the physiological characteristics of EEG signals, we have redesigned the spectral distribution of positional encoding:
\begin{equation}
PE(pos, 2i) = \sum_{k=4}^{45} \frac{\alpha_k}{\sqrt{k}} \sin\left( \frac{pos}{10000^{2i/d_k}} \right),
\label{eq:adaptive-encoding}
\end{equation}
where \( \alpha_k = \text{Softmax}(W_\omega k) \) denotes the learnable weight coefficient for each frequency band.

This design strictly restricts frequency components to the effective EEG frequency band (4-45Hz) while preserving the multi-scale characteristics of positional encoding. It also introduces learnable weight parameters \( \alpha_k \) to enable dynamic selection and enhancement of key frequency bands, thereby achieving an optimal balance between spectral adaptability and feature expression capability. This mechanism not only avoids interference from invalid high-frequency noise but also adaptively enhances the information representation of feature bands (such as \( \theta/\alpha \) bands) that are closely related to emotion recognition.

\subsubsection{2. Channel-Aware Dynamic Gating:}
To make rational use of spatial relationships, we constructed a channel attention gate and a feedback mechanism to achieve implicit modeling of channel relationships.

(1) Channel statistics extraction
\begin{equation}
z_c = \frac{1}{T} \sum_{t=1}^{T} X_{b,c,1,t}, \quad \forall c \in \{1, \ldots, C\}
\end{equation}

(2) Adaptive weight generation
\begin{equation}
\begin{split}
\alpha = \sigma(W_2 \delta(W_1 z + b_1) + b_2), \\
\quad W_1 \in \mathbb{R}^{C/r \times C}, W_2 \in \mathbb{R}^{C \times C/r}
\end{split}
\end{equation}

(3) Feature modulation
\begin{equation}
\hat{X}_{b,c,1,t} = X_{b,c,1,t} \cdot \alpha_c
\end{equation}

\subsubsection{3. Time-Frequency Attention:}

Multi-scale feature extraction is achieved by computing temporal and frequency domain attention in parallel:

\begin{equation}
\text{Head}_{\text{time}} = \text{Attention}(Q_{\text{time}}, K_{\text{time}}, V_{\text{time}})
\end{equation}
\begin{align}
\text{Head}_{\text{freq}} &= \text{DCT}^{-1}\bigl( A(Q_{\text{freq}}, K_{\text{freq}}, V_{\text{freq}}) \bigr) 
\end{align}
where $A = \text{Attention}\bigl( \text{DCT}(Q_{\text{freq}}), \text{DCT}(K_{\text{freq}}), \text{DCT}(V_{\text{freq}}) \bigr)$

The temporal domain head processes the original signal to preserve transient features, and the frequency domain head analyzes rhythmic activities and employs DCT to achieve efficient transformation.

The decoder reconstructs clean EEG signals via transposed convolution:
\begin{equation}
\hat{h}_{x,\text{clean}} = f_{\text{dec}}(h_{\text{denoise}}) + x_{\text{noisy}}
\end{equation}

Residual learning is adopted to preserve high-frequency components, where the information $h_{\text{denoise}}$ from the denoising path.

\subsection{Classify Module}
The classification head processes features from the classification path:  
\begin{equation}
y_{\text{pred}} = \text{Softmax}(c_{\text{cls}}(\text{Pool}(h_{\text{classify}})))
\end{equation}
where $h_{\text{classify}}$ in the features are enhanced by channel attention. We propose an adaptive Binary Cross-Entropy (BCE) loss based on the inverse square weighting of label frequencies:
\begin{equation}
\mathcal{L}_{\text{cls}} = -\sum_{c} w_c \bigl[ y_c \log p_c + (1 - y_c) \log (1 - p_c) \bigr]
\end{equation}
where the weight coefficient $w_c = 1 / \sqrt{f_c}, \, f_c$ is the class frequency.

\subsection{Dual-Path Feature Interaction with Feedback}
In traditional deep learning training, we observe that the gradient of the denoising task ($\nabla\mathcal{L}_{\mathrm{den}}$) conflicts with the gradient of the classification task ($\nabla\mathcal{L}_{\mathrm{cla}}$) in 30\% of the parameter space. During the feature disentanglement process, orthogonal constraint experiments revealed that noise-related features and emotional features in EEG signals overlap by 15\% in the frequency domain (specifically in the 4-8Hz theta band)\citep{kendall2018multi}.

To address these issues, we propose an interactive feedback module that balances both tasks and achieves optimal results through task synergy. This module enables collaborative optimization of denoising and classification via bidirectional feature flows, with mathematical formulation:

\begin{equation}
\left\{
\begin{aligned}
H_{\text{denoise}}^{(t)} &= \mathcal{F}_{\text{den}}\left(H_{\text{denoise}}^{(t-1)}, \Phi_{\text{cls}\to \text{den}}^{(t-1)}\right) \\
\Phi_{\text{den}\to \text{cls}}^{(t)} &= \mathcal{G}_{\text{fb}}\left(H_{\text{denoise}}^{(t)}\right) \\
H_{\text{classify}}^{(t)} &= \mathcal{F}_{\text{cls}}\left(H_{\text{classify}}^{(t-1)}, \Phi_{\text{den}\to \text{cls}}^{(t)}\right)
\end{aligned}
\right.
\end{equation}
where $t$ denotes iteration steps and $\Phi$ represents feature transformation functions.

Let the output of the valence / arousal probability of the classifier be \( p \in \mathbb{R}^2 \), and the feedback projection process is:

\subsubsection{1. Low-Dimensional Embedding:}
\begin{equation}
\mathbf{z} = \text{ReLU}(W_e \mathbf{x} + \mathbf{b}_e), 
\end{equation}
where $\mathbf{x} \in \mathbb{R}^D$ is the input high-dimensional feature vector, $W_e \in \mathbb{R}^{d \times D}$ is the embedding weight matrix, $\mathbf{b}_e$ is the bias term, $\mathbf{z} \in \mathbb{R}^d$ is the resulting low-dimensional representation, and $D \gg d$ ensures the reduction of dimensionality.

The high-dimensional data is projected into a low-dimensional latent space using a weight matrix \( W_e \) and the ReLU activation function, which retains key features while reducing computational complexity.

\subsubsection{2. Feature Enhancement:}
\begin{equation} 
X \leq X \odot \big(I + \sigma(W f + b)\big), \quad W \in \mathbb{R}^{d \times d}
\label{eq:feature_enhance}
\end{equation}
where $\odot$ denotes the Hadamard (element-wise) product, $\sigma(\cdot)$ denotes the Sigmoid activation function ($\sigma(x) = \frac{1}{1+e^{-x}}$), $W$ denotes the learnable weight matrix of size $d \times d$, $f$ denotes the input features from the previous layer ($f \in \mathbb{R}^d$), $b$ denotes the bias term ($b \in \mathbb{R}^d$), and $\mathbf{I}$ denotes the vector of ones for residual connection.

The feature interaction module processes inputs through two parallel paths, with the following specifications: 

(1) Denoising path: It employs position encoding based on EEG frequency band characteristics and Transformer layers to capture spectro-temporal patterns (within the 4-45Hz range). 

(2) Classification path: It utilizes a temporal attention mechanism to emphasize time segments associated with emotions.

The innovative feedback mechanism projects the classification results into the feature space and adjusts the encoder output: 
\begin{equation}
h_{\text{feedback}} = \text{Sigmoid}(W_f y_{\text{pred}} + b_f)
\end{equation}
where \( W_f \in \mathbb{R}^{d \times c} \), and in the dimensionality-reduced prediction, it is transformed into the feature dimension.

\subsubsection{3. Joint Training: }
The total loss combines the denoising Mean Squared Error (MSE) and the classification loss:
\begin{equation}
\mathcal{L}_{\text{total}} = \alpha \lVert x_{\text{clean}} - \hat{x}_{\text{clean}} \rVert_2^2 + (1 - \alpha) \mathcal{L}_{\text{cls}}
\end{equation}
The experimental setup balances task weights with \(\alpha = 0.6\). During training, a curriculum learning strategy is employed, gradually reducing the Signal-to-Noise Ratio (SNR) from 3 dB to -3 dB.

\section{Experiments}
\subsection{Datasets}
We evaluated FDC-Net on two public datasets: DEAP and DREAMER. DEAP induces emotions via music videos, collecting 32-channel EEG (per 10-20 system) and peripheral signals from 32 participants. EEG was sampled at 512 Hz with a 0.5-45 Hz band-pass filter, annotated with 4 dimensions (valence, arousal, etc.) using 1-9 continuous ratings. DREAMER gathers 14-channel EEG (Emotiv EPOC+) from 23 participants watching emotional videos, sampled at 128 Hz with a 0.1-45 Hz filter, and annotated with 3 dimensions (valence, arousal, and dominance) using 1-5 discrete ratings.
Experiments focused on valence and arousal, discretized into high/low binary classification tasks. Both datasets were expanded via 50\% overlapping 128-length sliding windows, yielding 160,000 and 55,062 EEG segments for DEAP and DREAMER, respectively.
 
\subsection{Noise Addition}
In this study, we employed a composite noise addition approach to simulate the EEG signals with artifacts in real-world scenarios \citep{lai2018artifacts}. Specifically, the noise signal was formed by mixing EMG and EOG signals at a 1:1 ratio. For the generation of EMG noise, one channel was randomly selected from the 2 EMG channels in the original data for sampling, and this process was repeated 32 times. The same strategy was applied to generate EOG noise from the 2 EOG channels. To ensure precise control over noise intensity, we dynamically adjusted the noise amplitude coefficient $\lambda_{\text{noise}}$ according to the preset SNR by calculating the ratio of the root mean square values of the original EEG signal and the mixed noise. Furthermore, to enhance the robustness of the model, a small amount of Gaussian white noise (with a standard deviation of 0.01) was introduced in addition to the bioelectrical noise. All noise injection processes were completed before sliding window segmentation. This processing method not only preserves the temporal integrity of EEG signals but also effectively simulates complex interference conditions in actual acquisition environments.

\begin{figure}
    \centering
    \begin{subfigure}[b]{0.48\linewidth} 
        \centering
        \includegraphics[width=\linewidth]{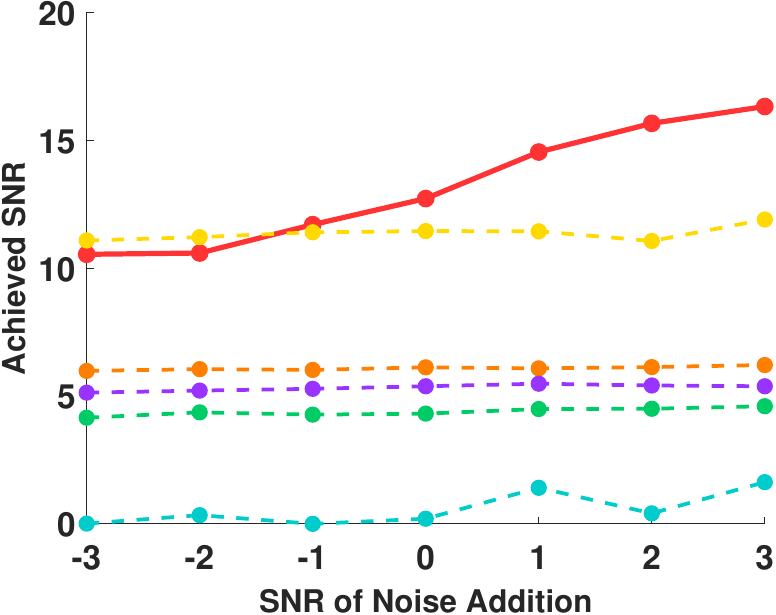}
        \caption{SNR}
        \label{snr_deap}
    \end{subfigure}
    \hfill
    \begin{subfigure}[b]{0.48\linewidth} 
        \centering
        \includegraphics[width=\linewidth]{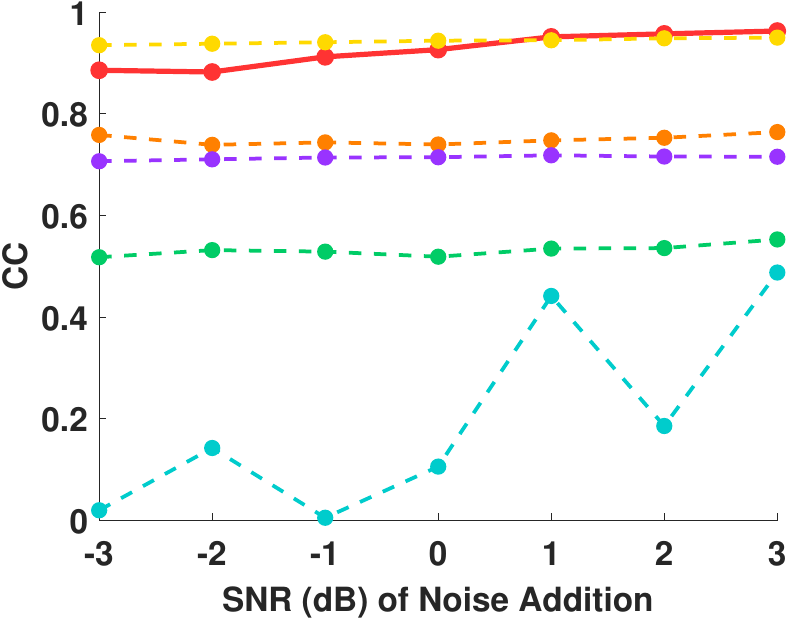}
        \caption{CC} 
        \label{cc_deap}
    \end{subfigure}
    
    \begin{subfigure}[t]{0.48\linewidth} 
        \centering
        \includegraphics[width=\linewidth]{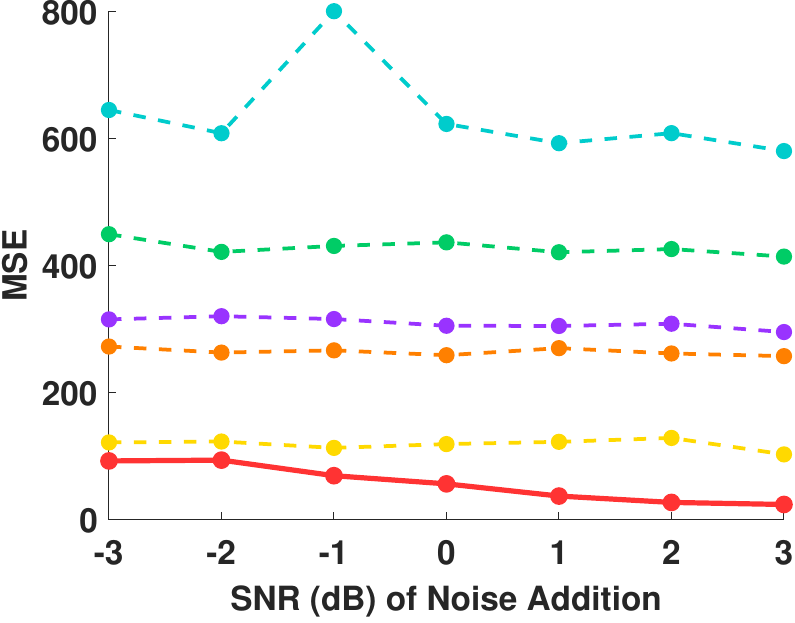} 
        \caption{MSE} 
        \label{mse_deap}
    \end{subfigure}
    \hfill
   \begin{subfigure}[t]{0.5\linewidth} 
        \centering
        \includegraphics[width=0.5\linewidth]{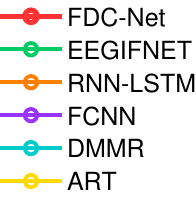} 
        \label{legend}
    \end{subfigure}
    
    \caption{Comparison of the denoising performance under varying signal-to-noise ratios on DEAP.}
    \label{fig:denoise_deap}
\end{figure}

\begin{figure}
    \centering
    \begin{subfigure}[b]{0.48\linewidth} 
        \centering
        \includegraphics[width=\linewidth]{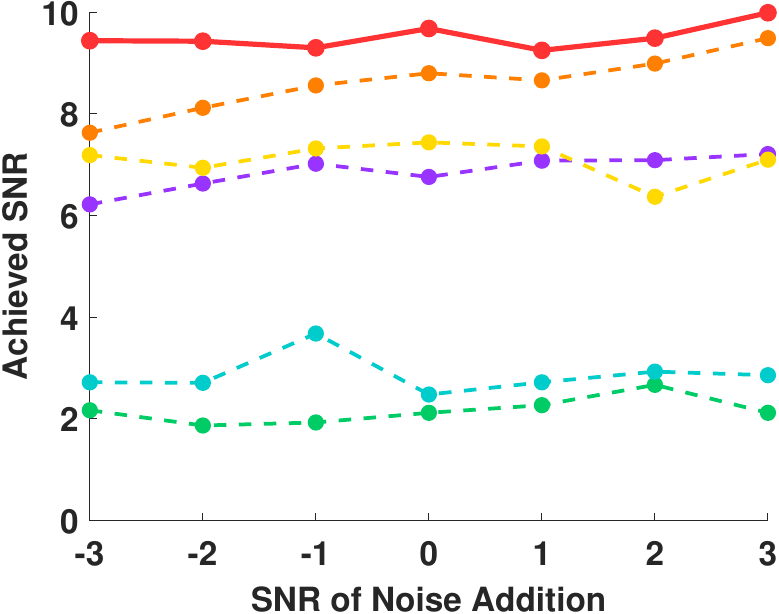}
        \caption{ SNR}
        \label{snr_dreamer}
    \end{subfigure}
    \hfill
    \begin{subfigure}[b]{0.48\linewidth} 
        \centering
        \includegraphics[width=\linewidth]{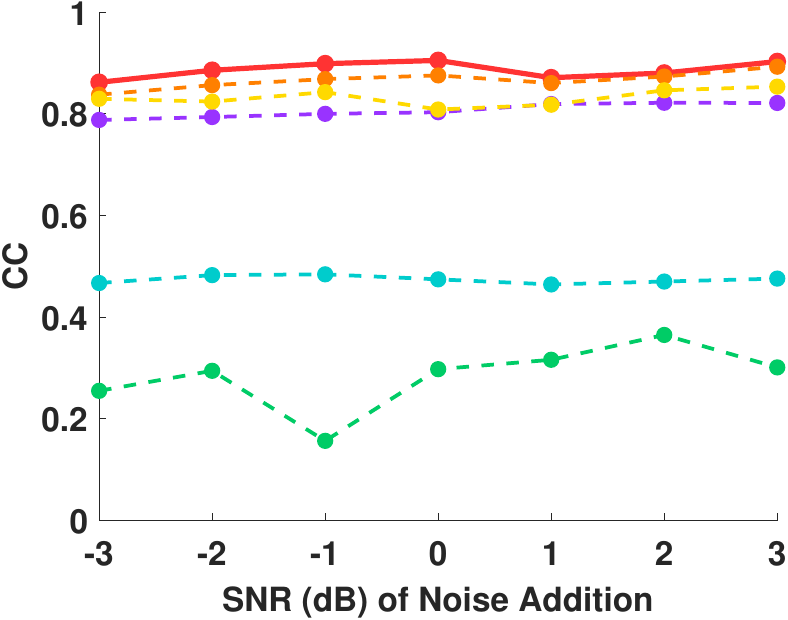}
        \caption{CC} 
        \label{cc_dreamer}
    \end{subfigure}
    
    \begin{subfigure}[t]{0.48\linewidth} 
        \centering
        \includegraphics[width=\linewidth]{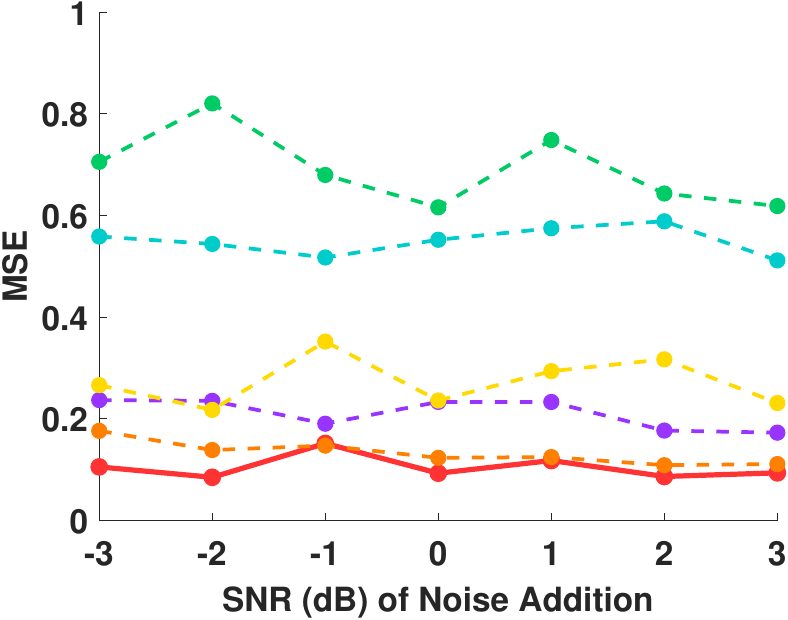} 
        \caption{MSE} 
        \label{mse_dreamer}
    \end{subfigure}
    \hfill
    \begin{subfigure}[t]{0.5\linewidth} 
        \centering
        \includegraphics[width=0.5\linewidth]{tuli_denoise_2.pdf} 
        \label{legend}
    \end{subfigure}
    
    \caption{Comparison of the denoising performance under varying signal-to-noise ratios on DREAMER.}
    \label{fig:denoise_dreamer}
\end{figure}

\begin{figure}
    \centering
    \begin{subfigure}[b]{0.42\linewidth} 
        \centering
        \includegraphics[width=\linewidth]{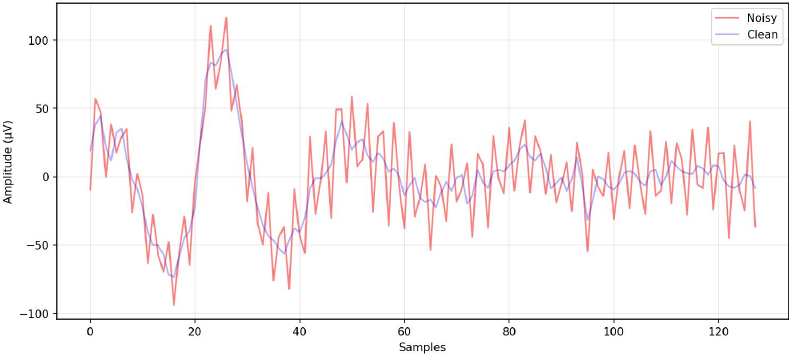}
        \caption{Noisy signals} 
        \label{before_deap}
    \end{subfigure}
    \begin{subfigure}[b]{0.42\linewidth} 
        \centering
        \includegraphics[width=\linewidth]{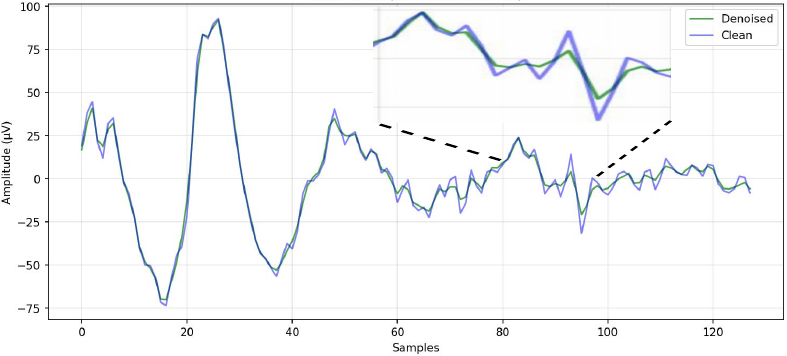}
        \caption{Denoised signals} 
        \label{after_deap}
    \end{subfigure}

    \begin{subfigure}[t]{0.42\linewidth} 
        \centering
        \includegraphics[width=\linewidth]{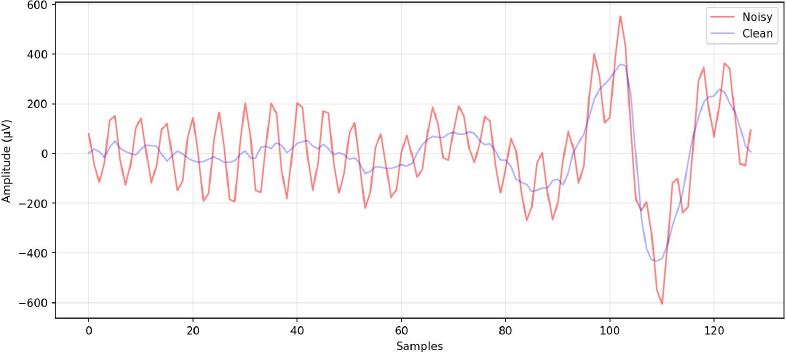}
        \caption{Noisy signals} 
        \label{before_dreamer}
    \end{subfigure}
    \begin{subfigure}[t]{0.42\linewidth} 
        \centering
        \includegraphics[width=\linewidth]{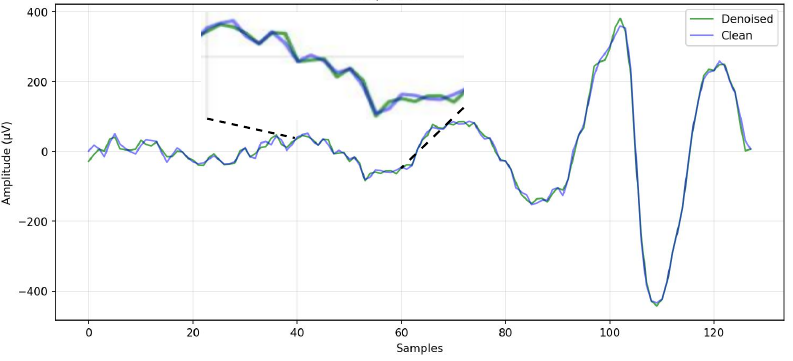}
        \caption{Denoised signals} 
        \label{after_dreamer}
    \end{subfigure}
    
    \caption{Comparison of noisy signals and denoised signals. The denoised signals obtained by DCF-Net. First and second rows represent DEAP and DREAMER, respectively.}
    \label{fig:denoise_wave}
\end{figure}

\begin{table}[htbp]
\centering
\label{tab:metrics}
\resizebox{\linewidth}{!}{%
\footnotesize  
\setlength{\tabcolsep}{3pt}  
\begin{tabular}{l *{6}{r}}  
\toprule
\multirow{2}{*}{Method} & \multicolumn{3}{c}{DEAP} & \multicolumn{3}{c}{DREAMER} \\
\cmidrule(lr){2-4} \cmidrule(lr){5-7}  
& MSE & CC/\% & SNR & MSE & CC/\% & SNR \\
\midrule
\textbf{FDC-Net (2025)} & \textbf{56.97} & \textbf{92.55} & \textbf{13.15} & \textbf{0.1049} & \textbf{87.08} & \textbf{9.51} \\
ART (2025)       & 118.52 & 94.31 & 11.36 & 0.2736 & 83.20 & 7.10  \\
DMMR (2024)      & 637.03 & 19.86 & 0.57 & 0.5498 & 47.43 & 2.87  \\
EEGIFNET (2023)  & 428.56 & 53.17 & 4.38 & 0.6909 & 28.40 & 2.16 \\
RNN-LSTM (2022)  & 264.25 & 74.95 & 6.08 & 0.1331 & 86.66 & 8.61 \\
FCNN (2021)      & 309.33 & 71.38 & 5.32 & 0.2115 & 80.70 & 6.86 \\
\bottomrule
\end{tabular}
}
\caption{Comparison of average results of different denoising methods.}
\label{tab:denoise_avg}
\end{table}

\begin{figure}
    \centering
     \begin{subfigure}[b]{1\linewidth} 
        \centering
        \includegraphics[width=\linewidth]{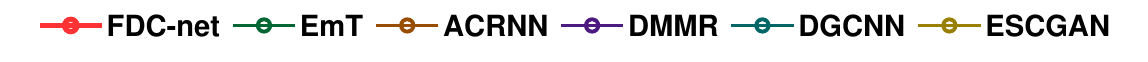}
        \label{tuli_classify}
    \end{subfigure}
    \begin{subfigure}[b]{0.45\linewidth} 
        \centering
        \includegraphics[width=\linewidth]{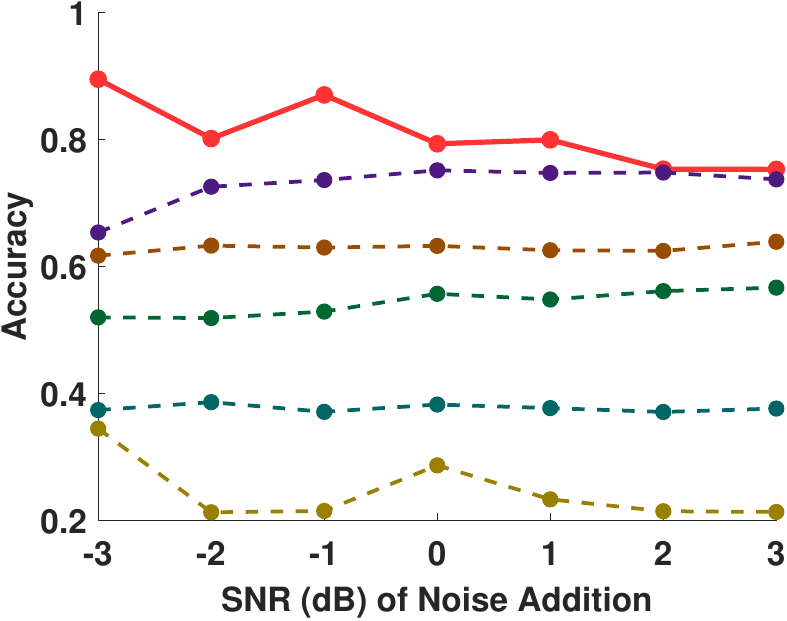}
        \caption{DEAP} 
        \label{accuracy_deap}
    \end{subfigure}
   \hfill
    \begin{subfigure}[b]{0.45\linewidth} 
        \centering
        \includegraphics[width=\linewidth]{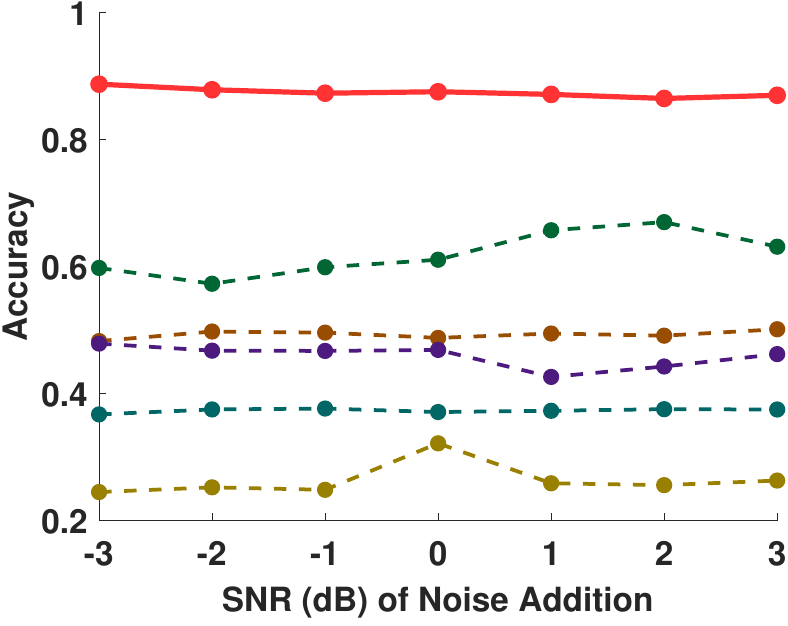}
        \caption{DREAMER} 
        \label{accuracy_dreamer}
    \end{subfigure}
    
    \caption{Comparison of the emotion recognition performance under varying signal-to-noise ratios.}
    \label{fig:classify}
\end{figure}

\begin{figure}
    \centering
    \begin{subfigure}[b]{0.45\linewidth} 
        \centering
        \includegraphics[width=\linewidth]{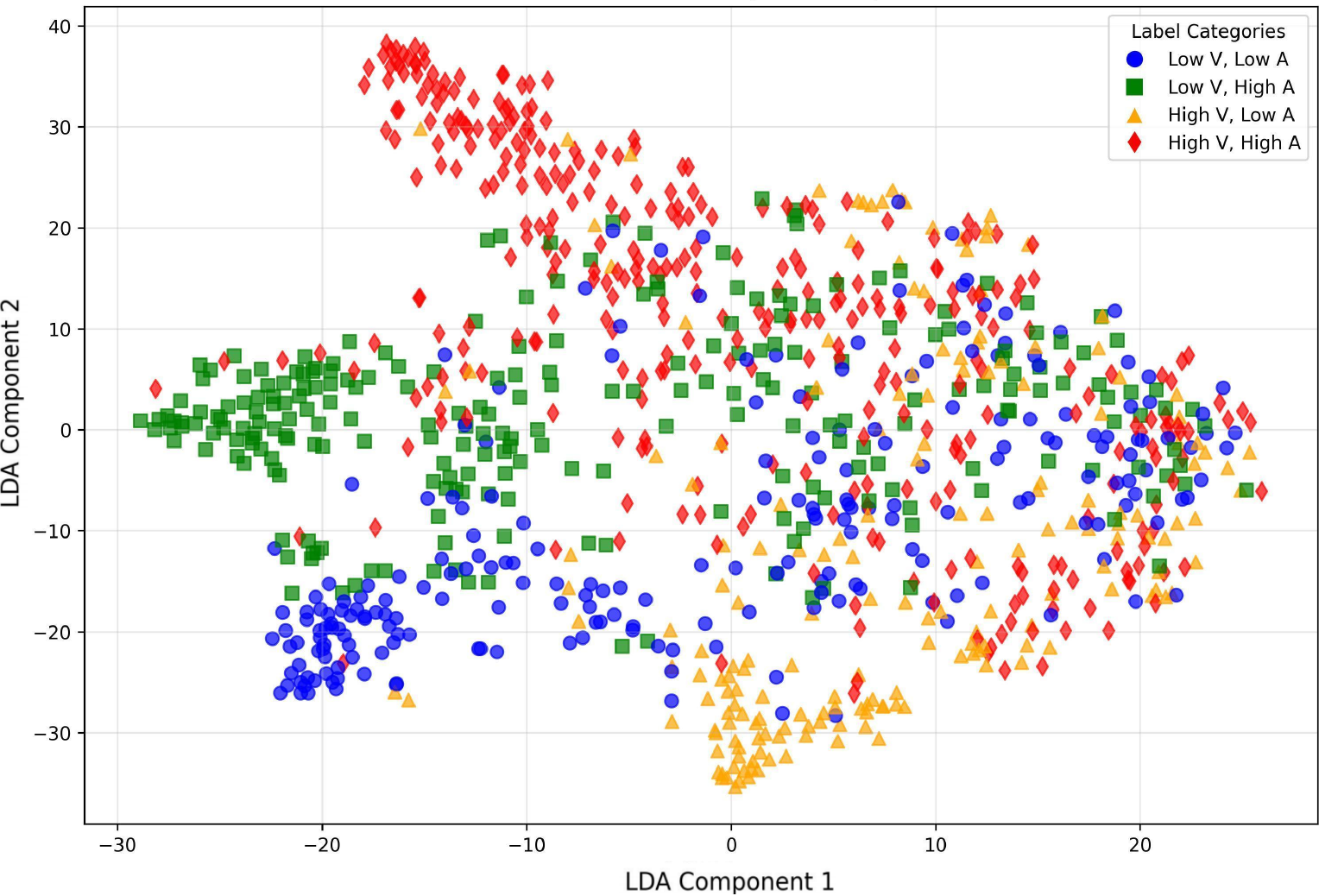}
        \caption{Before denoising} 
        \label{before_deap}
    \end{subfigure}
   \hfill
    \begin{subfigure}[b]{0.45\linewidth} 
        \centering
        \includegraphics[width=\linewidth]{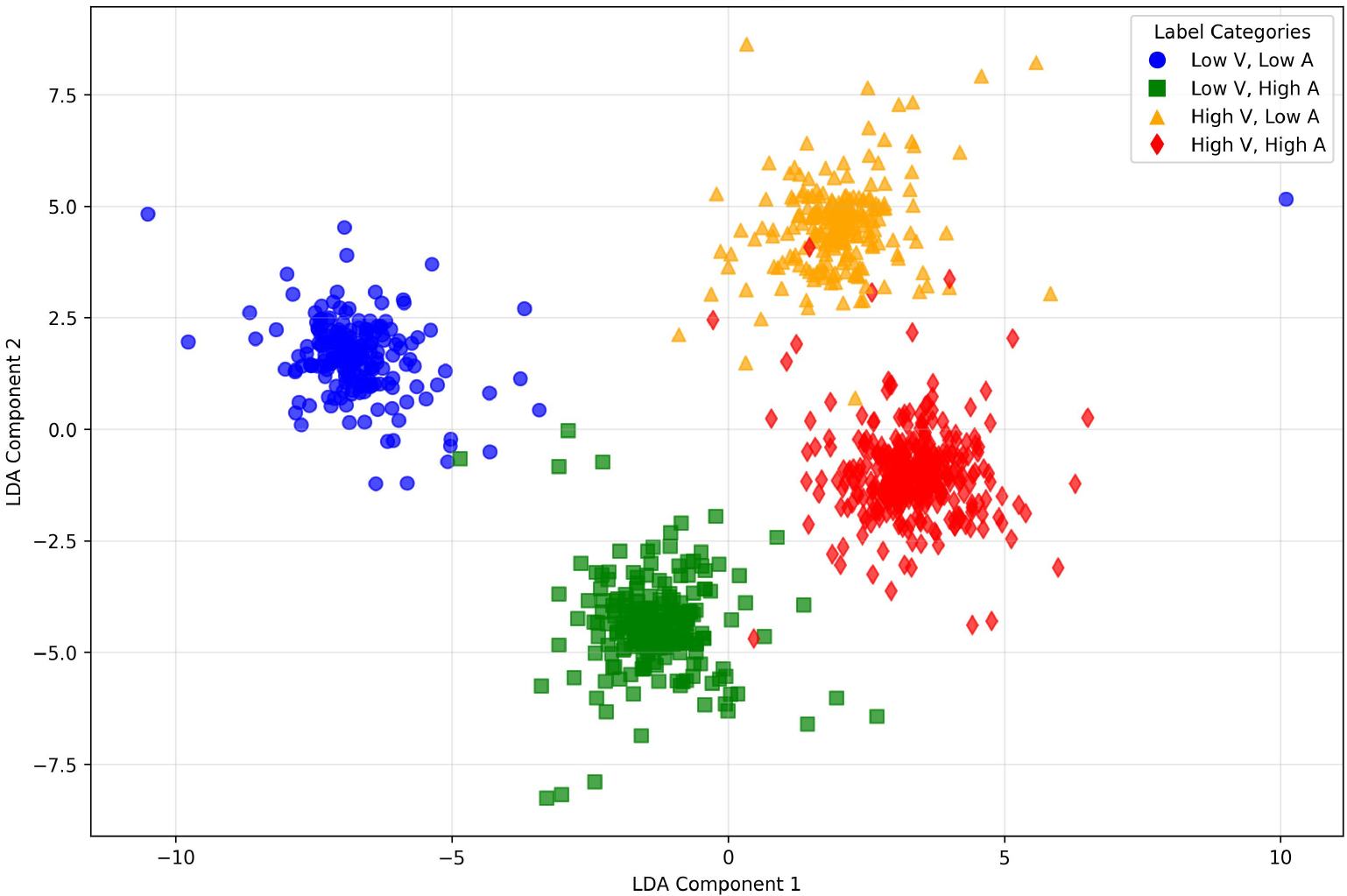}
        \caption{Before classification} 
        \label{after_deap}
    \end{subfigure}

    \begin{subfigure}[b]{0.45\linewidth} 
        \centering
        \includegraphics[width=\linewidth]{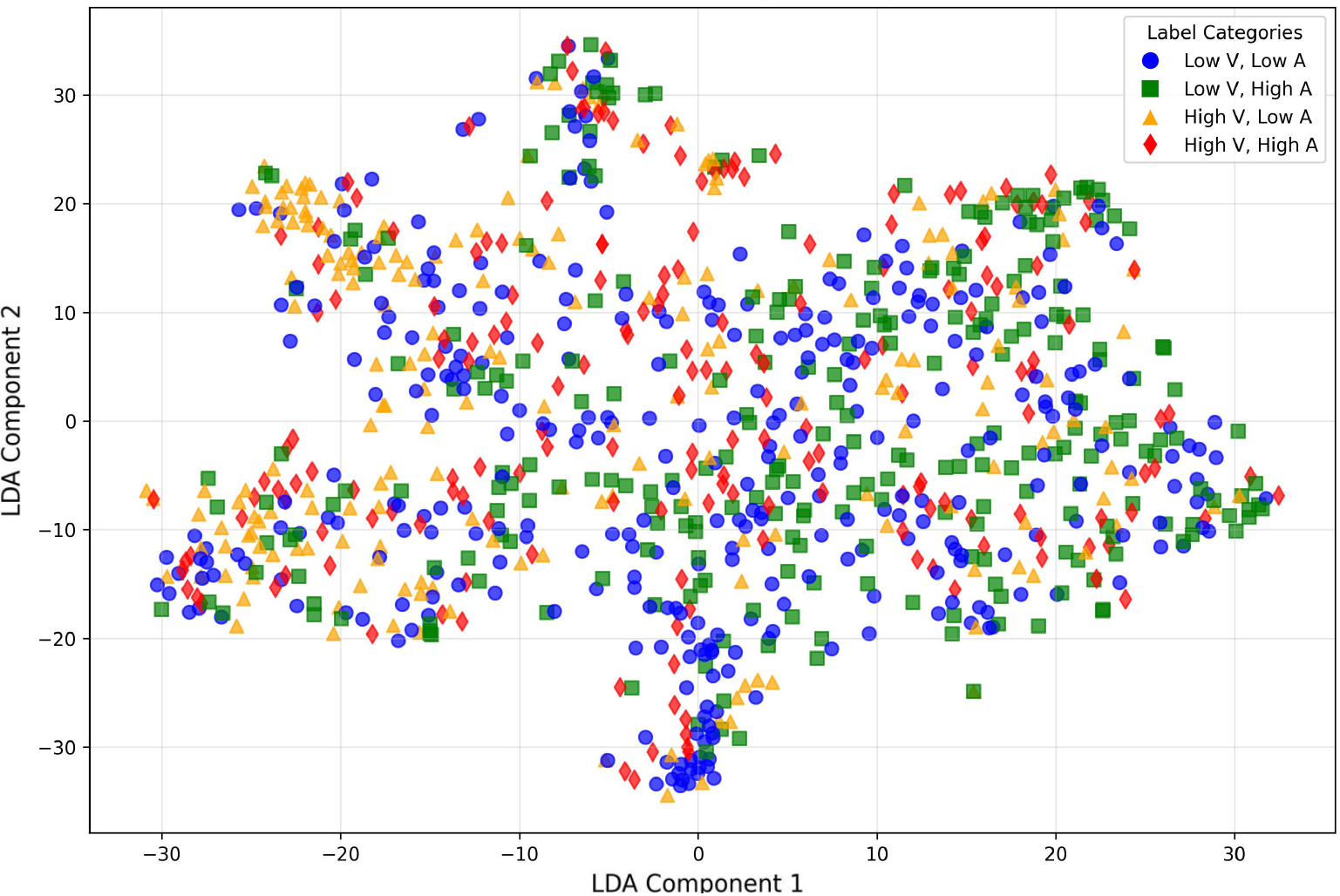}
        \caption{Before denoising} 
        \label{before_dreamer}
    \end{subfigure}
   \hfill
    \begin{subfigure}[b]{0.45\linewidth} 
        \centering
        \includegraphics[width=\linewidth]{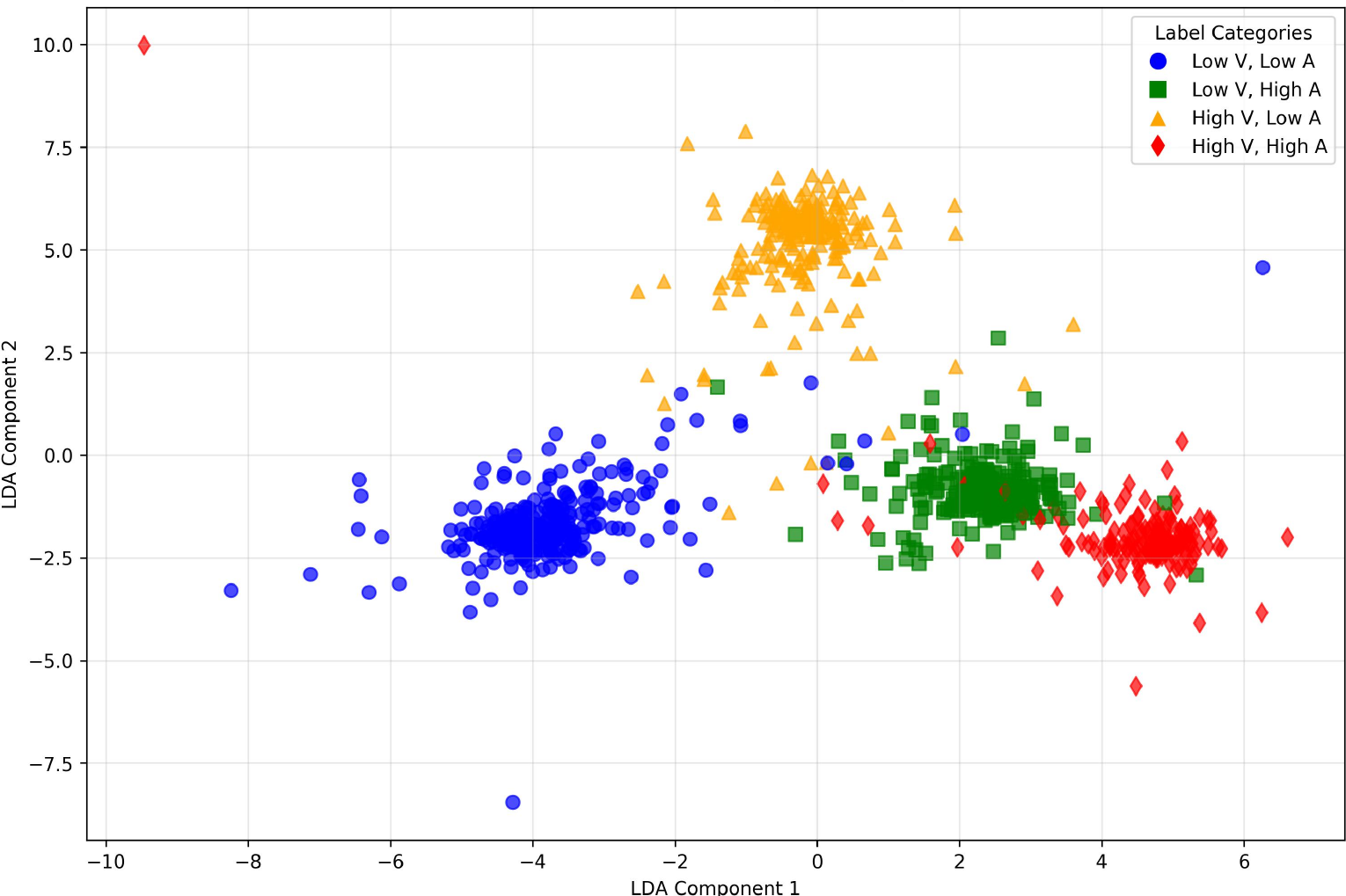}
        \caption{Before classification} 
        \label{after_dreamer}
    \end{subfigure}
    
    \caption{Feature visualization: On the left are the representations before denoising, and on the right are those the representations before classification. The top row presents results on the DEAP dataset, while the bottom row shows those on the DRAMER dataset.}
    \label{fig:classify_visual}
\end{figure}

\subsection{Implementation Details}
Experiments were performed on the DEAP and DREAMER datasets, with data randomly split into 80\% training and 20\% test sets, and results averaged over 10 repetitions. Training used the AdamW optimizer (initial learning rate 0.001, batch size 32). Evaluation metrics included SNR, CC, and MSE for denoising; for classification, 4-class accuracy was computed using combined valence and arousal labels. Metric trends under varying SNR conditions were recorded to analyze model robustness.  

\begin{table}[!t]
\centering
\small
\begin{tabular}{lcc}  
\toprule
Method & DEAP & DREAMER \\  
\midrule
\textbf{FDC-Net (2025)} & \textbf{0.8090} & \textbf{0.8737} \\  
EmT (2025) & 0.5427 & 0.6195 \\
ESC-GAN (2024) & 0.2458 & 0.2634 \\
DMMR (2024) & 0.7281 & 0.4588 \\
ACRNN (2020) & 0.6285 & 0.4927 \\
DGCNN (2018) & 0.3768 & 0.3730 \\
\bottomrule
\end{tabular}
\caption{Comparison of average results of different classification methods.}
\label{tab:classify_avg}
\end{table}

Model hyperparameters: input 32-channel EEG signals (128 time points, 1s); 128-dimensional hidden space (d\_model=128); Transformer with 2 encoding layers (8-head attention, feed-forward dimension 256); GELU activation, dropout 0.1 (to prevent overfitting); classifier: 2-layer fully connected network (hidden dim 64, output dim 2 for binary valence/arousal classification). Initial denoising-classification loss weight ratio: 0.6:0.4. All experiments were implemented in PyTorch on an NVIDIA RTX 5090 GPU. 

\subsection{Comparison Methods}
\subsubsection{1. Denoising comparative methods:}  
The denoising comparative methods include: \textbf{ART}\citep{chuang2025augmenting}; \textbf{DMMR}\citep{wang2024dmmr}; \textbf{EEGIFNet}\citep{10376353}; \textbf{RNN-LSTM}\citep{al2024rnn}; \textbf{FCNN}\citep{pratt2017fcnn}.  

\subsubsection{2. Classification comparative methods:}  
The classification comparative methods include: \textbf{EmT}\citep{ding2025emt}; \textbf{DMMR}\citep{wang2024dmmr}; \textbf{ESC-GAN}\citep{zhang2024beyond}; \textbf{ACRNN}\citep{tao2020eeg}; \textbf{DGCNN}\citep{song2018eeg}.

\subsection{Comparison with Denoising Methods}
As reflected in the trend analysis (Fig.~\ref{fig:denoise_deap} and~\ref{fig:denoise_dreamer}), across varying SNR levels of added noise on both DEAP and DREAMER datasets, the post-denoising SNR and CC curves of FDC-Net remain consistently high and exhibit a steady upward trend. The results indicates that FDC-Net maintains stable and effective noise suppression performance even under fluctuating noise intensities, outperforming comparative methods such as EEGINET and RNN\textunderscore LSTM in enhancing signal quality. 

Table~\ref{tab:denoise_avg} shows average result comparison of different denoising methods. As shown in Table~\ref{tab:denoise_avg}, in the average results across multiple SNR conditions, FDC-Net outperforms others in all key metrics. It achieves higher post-denoising SNR and CC, along with lower MSE, which further confirms its robust noise suppression capability. Furthermore, the waveform comparison in (Fig.~\ref{fig:denoise_wave}) intuitively illustrates the superiority of FDC-Net. In Fig.~\ref{fig:denoise_wave}, Fig.~\ref{fig:denoise_wave}(a) and Fig.~\ref{fig:denoise_wave} (b) are selected from P3 electrode data of the DEAP dataset, and Fig.~\ref{fig:denoise_wave}(c) and Fig.~\ref{fig:denoise_wave}(d) are selected from Fp1 electrode data of the DREAMER dataset. The method effectively preserves valuable physiological signal components while removing noise, yielding cleaner and more interpretable waveforms. 

\subsection{Comparison with Classification Methods}
As shown in the trend analysis (Fig.~\ref{fig:classify}), whether under different SNR conditions of added noise or in the comparison between the two datasets, its classification accuracy is higher than that of comparative methods, especially at lower SNR. The DMMR method has a certain noise suppression effect, and it can be seen from the results that the accuracy increases with the increase of initial SNR on DEAP. The accuracy of our method decreases to a certain extent as SNR increases, which implies that in our noisy data, partial noise still contains information useful for the downstream emotion classification task, and thus, denoising is not necessarily better when done "cleaner" to an excessive extent.

In the average classification results (Table~\ref{tab:classify_avg}), FDC-Net maintains higher accuracy, which can be attributed to its synergistic design: effective denoising provides cleaner input features for classification, while the classification task guides denoising to retain emotion-related information. Notably, the visualization of representations before and after classification (Fig.~\ref{fig:classify_visual}) clearly shows that FDC-Net generates more discriminative feature distributions, with distinct clustering of high/low valence/arousal classes, further validating its superiority in capturing emotion-related patterns.

\begin{table}[htbp]
  \centering
  \small
  \begin{tabular}{lcccc}
    \toprule
    & \multicolumn{2}{c}{DEAP} & \multicolumn{2}{c}{DREAMER} \\
    \cmidrule(lr){2-3} \cmidrule(lr){4-5}
    Method & Memory/MB & Time & Memory/MB & Time \\
    \midrule
    \textbf{FDC-Net} & \textbf{3.40} & \textbf{2.13} & \textbf{2.38} & \textbf{1.12} \\
    EEGIFNET & 45.82 & 11.76 & 20.14 & 5.84 \\
    RNN\_LSTM & 0.19 & 3.57 & 0.19 & 0.68 \\
    FCNN & 0.25 & 2.65 & 0.25 & 0.51 \\
    ART & 2.45 & 7.38 & 0.69 & 1.52 \\
    DMMR & 0.27 & 2.03 & 0.25 & 0.49 \\
    EmT & 3.16 & 2.91 & 2.03 & 0.95 \\
    ACRNN & 0.68 & 2.13 & 0.56 & 0.70 \\
    DGCNN & 0.01 & 3.67 & 0.00 & 0.61 \\
    ESC-GAN & 5.11 & 3.28 & 4.99 & 1.08 \\
    \bottomrule
  \end{tabular}
  \caption{Comparison of test time (seconds) and memory among different methods.}
  \label{tab:time_memory}
\end{table}
\subsection{Running Time and Parameter Size Analysis}
Table~\ref{tab:time_memory} compares FDC-Net with other methods in running time and memory. In terms of computational efficiency, FDC-Net exhibits a trade-off between parameter size and running time. Due to the integration of a Transformer architecture to model complex spatiotemporal dependencies, its parameter size is relatively larger compared to lightweight methods (e.g., FCNN), which is a result of the Transformer’s inherent need for sufficient capacity to capture long-range correlations.

However, FDC-Net still maintains fast running speed. This is mainly due to the optimized design of its internal modules: the cross-modal interaction and feedback mechanisms avoid redundant computations, and the streamlined Transformer structure balances expressive power and efficiency. As shown in Table~\ref{tab:time_memory}, despite the larger parameter size, the running time of FDC-Net is comparable to or even faster than some comparative methods with similar performance, making it suitable for practical application scenarios requiring real-time processing.

\begin{table}[H]
\centering
\small
\setlength{\tabcolsep}{3pt}  
\begin{tabular}{llcccc}
\toprule
Dataset & Method & MSE & CC/\% & SNR & Accuracy \\
\midrule
\multirow{6}{*}{DEAP} 
& \textbf{FDC-Net} & \textbf{56.97} & \textbf{92.95} & \textbf{13.75} & \textbf{0.8090} \\
& abl\_cross & 58.77 & 92.46 & 13.11 & 0.6913 \\
& abl\_denoise & - & - & - & 0.7463 \\
& abl\_classify & 55.00 & 92.69 & 13.24 & - \\
& abl\_feedback & 64.71 & 89.69 & 12.31 & 0.7946 \\
& abl\_EEGSPTrans & 62.62 & 90.82 & 12.31 & 0.7977 \\
\midrule
\multirow{6}{*}{DREAMER} 
& \textbf{FDC-Net} & \textbf{0.1049} & \textbf{88.67} & \textbf{9.51} & \textbf{0.8737} \\
& abl\_cross & 0.1911 & 87.39 & 8.08 & 0.8625 \\
& abl\_denoise & - & - & - & 0.6918 \\
& abl\_classify & 0.3329 & 87.66 & 9.49 & - \\
& abl\_feedback & 0.1648 & 88.42 & 8.36 & 0.8627 \\
& abl\_EEGSPTrans & 0.1818 & 83.44 & 8.06 & 0.8632 \\
\bottomrule
\end{tabular}
\caption{Ablation study results, where abl\_EEGSPTrans represents abl\_EEGSPTransformer.}
\label{tab:ablation}
\end{table}
    
\subsection{Ablation Studies}
To analyze the effectiveness of FDC-Net and collaborative architecture, we conducted ablation experiments on two datasets, and the results are shown in Table 5. Table~\ref{tab:ablation} presents the results of the ablation experiments. In the ablation experiments on the DEAP and DREAMER datasets, “Origin” (without ablation) serves as the performance baseline. When comparing the indicators after the ablation of modules such as “abl\_cross” and “abl\_denoise”, there is a downward trend to varying degrees. This indicates that these modules play a positive role in maintaining data quality and model accuracy. In particular, the effects of the cross module and the EEGSPTrans module are more significant. 

\section{Conclusion and Future Work}
For EEG-based emotion recognition, FDC-Net addresses noise removal and emotion recognition task collaboration via deep-coupled denoising-classification and dynamic collaboration. Bidirectional gradient optimization avoids cascaded error accumulation, and a gated Transformer with learnable encoding enables balanced synergy. The experimental results under heavy noise show superior performance, validating enhanced denoising/classification and offering a novel physiological signal processing paradigm.

In future work, we will further simplify the complexity of our model, and explore lightweight attention mechanisms and knowledge distillation to balance efficiency and performance. Additionally, we will extend FDC-Net to other physiological signals. Finally, we aim to integrate the framework with more EEG downstream tasks, such as clinical mood disorder diagnosis and cognitive load prediction, to amplify its practical utility.

\bibliography{aaai2026}

\end{document}